\documentclass[aps,prb,twocolumn,superscriptaddress,showpacs,floatfix]{revtex4}
\usepackage{epstopdf}
\usepackage{graphicx}
\usepackage{dcolumn}
\usepackage{bm}
\usepackage{amsfonts}
\usepackage{amsmath}
\usepackage{color}

\begin{document}
\renewcommand{\thefigure}{\arabic{figure}}
\def\be{\begin{equation}}
\def\ee{\end{equation}}
\def\ber{\begin{eqnarray}}
\def\eer{\end{eqnarray}}

\renewcommand{\arraystretch}{1.5}

\newcommand{\h}[1]{{\hat {#1}}}
\newcommand{\hdg}[1]{{\hat {#1}^\dagger}}
\newcommand{\bra}[1]{\left\langle{#1}\right|}
\newcommand{\ket}[1]{\left|{#1}\right\rangle}

\title{Unconventional proximity-induced superconductivity in bilayer systems}

\date{\today}

\author{Fariborz Parhizgar}\email{fariborz.parhizgar@ipm.ir}

\affiliation{School of Physics, Institute for Research in Fundamental Sciences (IPM), Tehran 19395-5531, Iran}
\affiliation{Department of Physics and Astronomy, Uppsala University, P.O. Box 516, S-751 20 Uppsala, Sweden}
\author{Annica M. Black-Schaffer}\email{annica.black-schaffer@physics.uu.se}
\affiliation{Department of Physics and Astronomy, Uppsala University, P.O. Box 516, S-751 20 Uppsala, Sweden}

\begin{abstract}
We study the proximity-induced superconducting state in a general bilayer -- conventional $s$-wave superconductor hybrid structure. For the bilayer we include a general parabolic dispersion, Rashba spin-orbit coupling, and finite layer tunneling as well as the possibility to apply a bias potential and a magnetic Zeeman field, in order to address experimentally relevant bilayer systems, ranging from topological insulator thin films to generic double quantum well systems.
By extracting the proximity-induced anomalous Green's function in the bilayer we show on a very rich structure for the superconducting pairing, including different spin states and odd-frequency pairing.
Equal-spin spin-triplet $p_x\pm ip_y$-wave pairing is induced in both layers in the presence of a finite spin-orbit coupling and opposite-spin spin-triplet $s$-wave pairing with odd-frequency dependence appears for an applied magnetic Zeeman field. Finite interlayer pairing is also generally present in the bilayer. The interlayer pairing can be either even or odd in the layer index, with a complete reciprocity between parity in frequency and in layer index.
We also find that a bilayer offers the possibility of sign reversal of the superconducting order parameters, both between the two layers and between multiple Fermi surfaces.
\end{abstract}

\maketitle

%
\section{Introduction}
\label{sec:intro}
The superconducting proximity effect offers a unique possibility to study superconductivity in combination with other exotic materials properties. By depositing a non-superconducting material (N) on top of a known superconductor (S), superconductivity  leaks into the N material and the resulting superconducting state combines the original properties of N with those of superconductivity.

One recent notable example of proximity-induced superconductivity is the measuring of a finite supercurrent in graphene by creating a SNS, or Josephson, junction in graphene.\cite{Heersche07, Miao07, Shailos07, Du08} Among the intriguing properties of graphene, the linear Dirac spectrum offers the possibility of specular Andreev reflection.\cite{Beenakker06}
Another contemporary proximity system is created by depositing a topological insulator (TI) on top of a superconductor.\cite{Fu08, Sacepe11, Veldhorst12, Zareapour12, Wang13}  A TI is insulating in the bulk and, due to the non-trivial topology of its band structure, it has a single conducting Dirac cone surfaces state (or an odd number).\cite{Kane05b, Bernevig06, Konig07, Fu07, Moore07, Hasan10, Qi11} The surface state has its spin locked to the momentum described by a two-dimensional (2D) Dirac Hamiltonian.\cite{Hsieh08, Hsieh09, DMreview}
Combining the TI surface state with superconductivity has been shown to result in an effective spinless chiral $p+ip'$-wave superconductor.\cite{Fu08} The main allure of such superconducting pairing is that it hosts Majorana fermions in vortex cores and Josephson junctions.\cite{Read00, Fu08, Alicea12, Beenakker13} The Majorana fermion is its own antiparticle and obeys non-Abelian statistics in 2D, which supports fault-tolerant quantum computation.\cite{Nayak08} 

As evident from both graphene and TIs, proximity-induced superconductivity can result in very interesting phenomena when the normal material is effectively 2D. Introducing a discrete third dimension in the normal system results in a bilayer structure in proximity-contact to the external superconductor. The additional layer degree of freedom naturally offers the possibility for yet more novel behavior. Among bilayer systems currently actively studied are those naturally formed such as bilayer graphene,\cite{Castro08, McCann13} where the two layers can also partly decouple due to stacking faults present when grown on a silicon carbide substrate.\cite{Sadowski06,Wu07, Varchon07, Lopes07, Hass08} Recently, superconductivity in bilayer graphene has been predicted to both give rise to retro-reflection of electrons \cite{Ang12} and be a playground for exotic spin-triplet $s$-wave interlayer pairing.\cite{Hosseini12} Also, ABC stacked multilayer graphene has been shown to host zero-energy surface states, recently predicted to easily become superconducting.\cite{Kopnin11, Munoz13}
TI thin films is another actively studied bilayer system. For film thicknesses less than six quintuple layers of the TI Bi$_2$Se$_3$ the surface states on the opposite sides of the film hybridize and create a finite energy gap.\cite{Zhang10, Lu10, Shan10} Proximity-induced superconductivity has recently been experimentally demonstrated in Bi$_2$Se$_3$ TI thin films by growing the TI on a conventional NbSe$_2$ superconductor.\cite{Wang12} A superconducting energy gap was found also on the surface away from the superconductor interface, with a diminishing size with increasing film thickness.
More generally, these and other bilayers systems can be seen as material specializations of the double layer formation of a 2D electronic gas in semiconductor heterostructures, creating an effective double quantum well system.\cite{Ferreira97, Hasbun03}
In this work we provide a theoretical study of the proximity-induced superconducting state in a general bilayer -- superconductor hybrid structure, with results applicable to bilayer systems ranging from TI thin films to generic double quantum well systems.

In terms of classifying the superconducting state in a general bilayer system, we first of all conclude that the layer degree of freedom adds another symmetry. The superconducting pair amplitude, being the wave function of the Cooper pairs, needs to obey Fermi-Dirac statistics. Thus a spin-singlet pair usually has even spatial parity ($P$) such as $s$- and $d$-wave, whereas spin-triplet pairing gives odd spatial parity ($p$-wave).
In addition, the superconducting pairing amplitude can also be even or odd in time ($T$), or equivalently frequency,\cite{Berezinskii74, Balatsky92} with the relation $PT = +1(-1)$ for spin-singlet (triplet) pairing.
Some theoretical proposals exist for intrinsic odd-frequency bulk superconductors,\cite{Kirkpatrick91, Balatsky92, Abrahams95} but odd-frequency superconductivity has mostly been associated with interfaces and surfaces.\cite{Tanaka12}
For example, superconductor -- ferromagnet interfaces have been shown to host odd-frequency spin-triplet $s$-wave pairing responsible for a long-range proximity effect into the ferromagnet.\cite{Bergeret01,Bergeret05}
More recently, odd-frequency spin-triplet $s$-wave pairing has also been shown to appear in TIs without invoking any interface phenomenon, only requiring a superconducting state with a spatially varying amplitude or phase,\cite{Black-Schaffer12oddw} or if in proximity-contact with a spin-triplet $p$-wave superconductor.\cite{Black-Schaffer13TISC}
Finally, an additional layer degree of freedom ($L$) gives rise to $PTL = +1(-1)$ for spin-singlet (triplet) pairing. Superconducting pairing within the same layer, i.e.~intralayer pairing, is naturally always even in $L$, but if the two electrons forming a Cooper pair belongs to different layers both even ($L = 1$) and odd ($L = -1$) interlayer pairing can be present. In fact, the layer index acts symmetry-wise, as we will show, in the same way as the band (or orbital) index in multiband superconductors.\cite{Black-Schaffer13multi}

\begin{figure}
\includegraphics[width=0.49\linewidth]{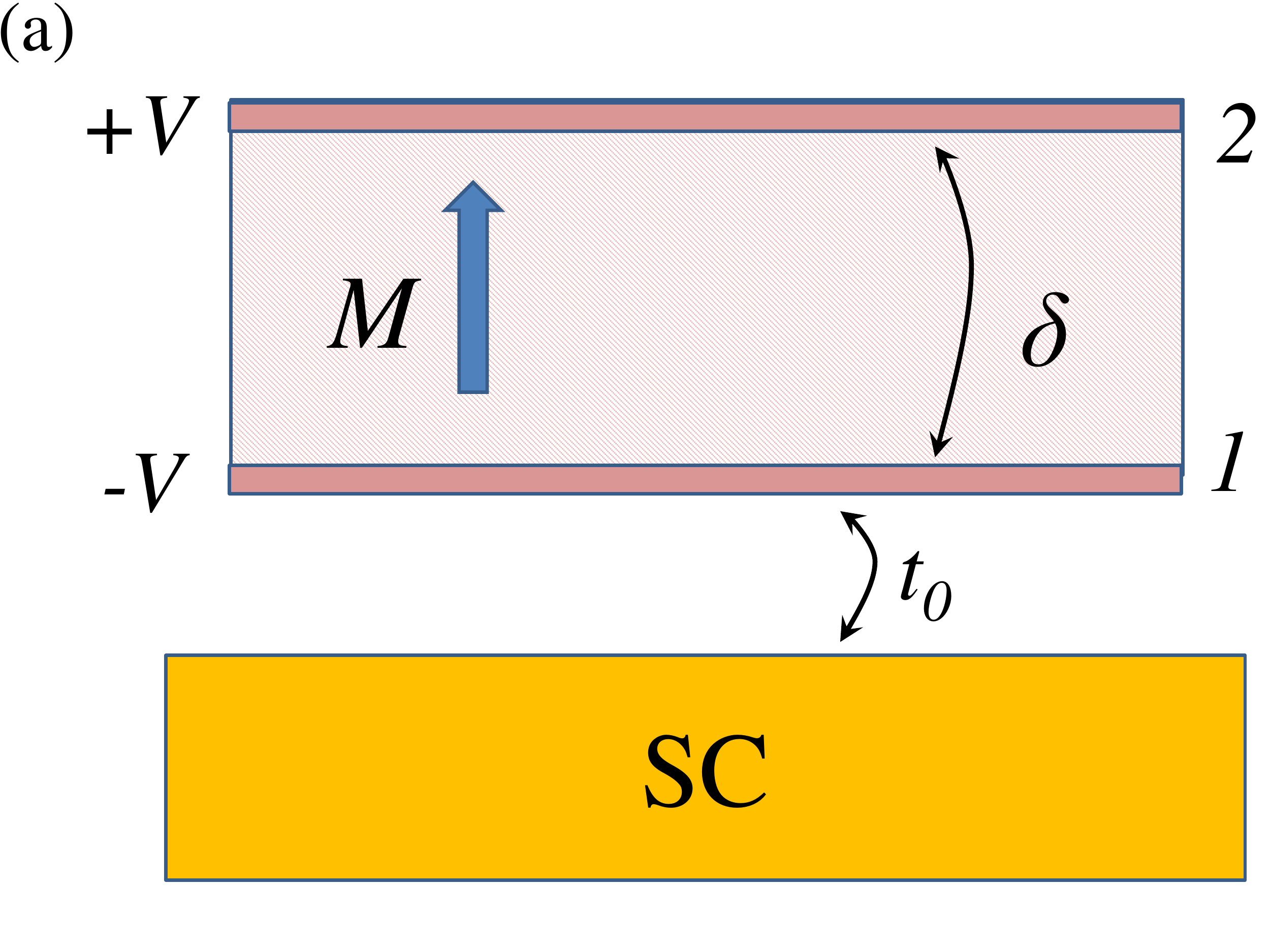}
\includegraphics[width=0.49\linewidth]{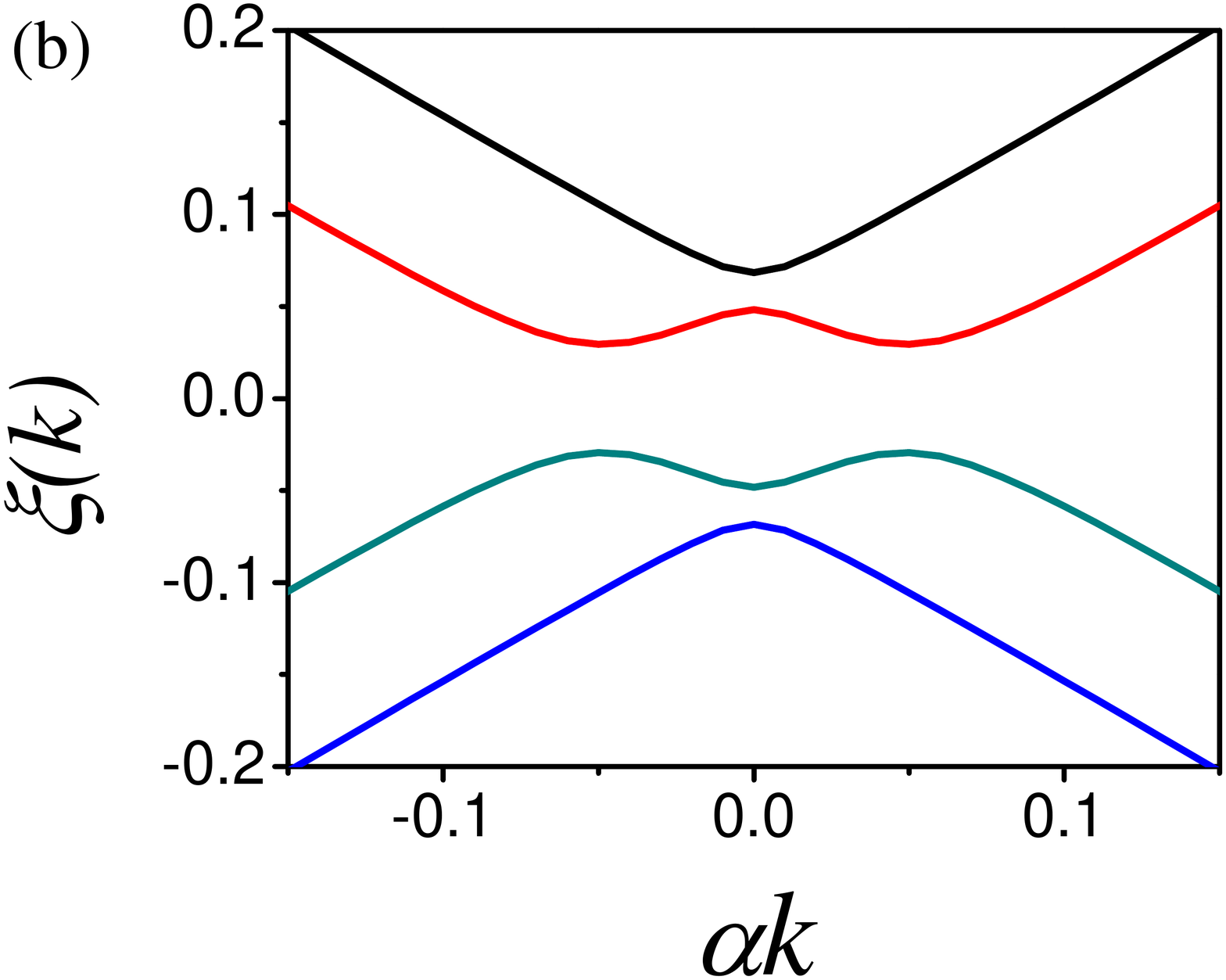}
\caption{\label{fig:schem}(Color online.) (a) Bilayer system (pink) in proximity to a spin-singlet $s$-wave superconductor (yellow). The superconductor is connected by a single-particle tunneling amplitude $t_0$ to the bottom layer (1) of the bilayer system whereas the two bilayers are coupled by an interlayer tunneling $\delta$. A finite magnetization $M$ in the $z$-direction and a bias voltage $2V$ can also be applied to the bilayer.
(b) Energy dispersion (eV) of a TI thin film as function of the momentum scaled by the spin-orbit interaction $\alpha$ (eV) for $\delta = 0.03$, $M = 0.01$, and $V = 0.05$~eV.
}
\end{figure}

To gain a more detailed understanding of the superconducting proximity effect in bilayer systems, we study a general bilayer in proximity to a conventional spin-singlet $s$-wave superconductor, as schematically illustrated in Fig.~\ref{fig:schem}(a).
In order to study a generic bilayer we include both a quadratic energy dispersion and a Rasbha spin-orbit coupling as intrinsic properties of the bilayer, as well as assume that both an electric bias voltage between the layers and a magnetic Zeeman field can be applied to the bilayer. A double quantum well system is realized by ignoring the Rashba coupling, whereas a TI thin film is produced if the quadratic dispersion is neglected.\cite{Zhang10}
By solving the resulting Bogoliubov-de Gennes equation for the full hybrid system we find a finite superconducting gap with accompanied coherence peaks in both layers, but with a smaller gap in the layer farthest from the superconductor.
To more explicitly study the induced superconducting state we also calculate the anomalous Green's function in the bilayer. We find a very rich structure of different superconducting pairing amplitudes, where, e.g., a finite Rashba spin-orbit coupling induces equal-spin spin-triplet $p+ip'$-wave pairing and a finite magnetization gives rise to opposite-spin spin-triplet $s$-wave pairing with odd-frequency dependence.
Moreover, we find that both even and odd interlayer pairing, with the same spin and spatial symmetry as the intralayer pairing, is generally present. Since intralayer pairing is necessarily even in layer index, this means that the odd-interlayer pairing has the {\it opposite} frequency symmetry as that of the intralayer (and even-interlayer) pairing. In other words, we find a complete reciprocity between oddness in layer index and in frequency.
Furthermore, we find that the superconducting order parameter often reverses it sign between the two layers. In bilayer systems with two distinct Fermi surfaces, such as a biased TI thin film, we also find that the superconducting order parameter often reverses it sign between the two Fermi surfaces. Such sign-switching is present in the $s_{\pm}$-wave iron-pnictide superconductors\cite{Mazin08, Kuroki08, Mazin10} and have been shown to give rise to time-reversal invariant topological superconducting states,\cite{Deng12,Zhang13, Keselman13, Flensberg14} but here only a conventional $s$-wave superconductor is needed. 

The structure of the remainder of the article is as follows. In section \ref{sec:model} we introduce the model and how to extract the proximity-induced Green's functions in the bilayer system. Then in Section \ref{sec:disp} we calculate the band structure and local density of states (LDOS) in each layer.  In Section \ref{sec:Green} we focus on the anomalous Green's function, breaking down the results into several limits before presenting the results for a generic bilayer system. In Section \ref{sec:OP} we study the superconducting order parameter and analyze the conditions for sign-switches in the order parameter. Finally, in Section \ref{sec:sum} we summarize our results and provide some final remarks.

\section{Method} \label{sec:model}
The Hamiltonian for a bilayer system, including a Rashba spin-orbit coupling $\alpha$ and in the presence of a Zeeman field $M$ and a bias voltage $2V$ between the two layers, can be written as
\begin{align}\label{eq:H TI}
\hat{{\cal H}}_{BL}=\sum_k \{ K \sigma_0 \otimes \tau_0 +
 \alpha(\textbf{k} \times \sigma)\otimes \tau_z +\nonumber\\
  M \sigma_z \otimes \tau_0 +
V \sigma_0 \otimes \tau_z +\delta \sigma_0 \otimes \tau_x \},
\end{align}
where $\sigma$ and $\tau$ are Pauli matrices in spin and layer space, respectively. Here $K=\beta k^2 +\mu$, where $\beta$ denotes the coefficient for the ordinary kinetic energy $\hbar^2/(2m^*)$ and
$\mu$ is the chemical potential of bilayer, $\textbf{k}=(k_x,k_y,0)$ since the layers are 2D, and $\delta$ is the single-particle spin-independent hybridization between the two layers. We choose different directions for the Rashba term in the two layers, such that $\hat{{\cal H}}_{BL}$ describes a TI thin film in the $\beta\rightarrow 0$ limit.

Superconductivity is induced in the bilayer structure by contacting a conventional spin-singlet $s$-wave superconductor to the bottom surface, as illustrated in Fig.~\ref{fig:schem} (a). The Hamiltonian for the superconductor can be written as
\begin{align}\label{eq:H SC}
\hat{{\cal H}}_{SC}=\sum_{k,\lambda} \epsilon_k d^\dagger_{k,\lambda}d_{k,\lambda}+\sum_k \Delta_0
d^\dagger_{k,\uparrow}d^\dagger_{-k,\downarrow}+{\rm H.c.},
\end{align}
where $d$ ($d^\dagger$) is the annihilation (creation) operator in the superconductor. Here $\lambda$ labels the spin degree of freedom and $\Delta_0$ is the superconducting pairing order parameter. The energy dispersion is given by $\epsilon_k$, which we assume to be of the generic form $\hbar^2 k^2/(2m_{sc})+\mu_{sc}$.

We can calculate the Green's function of the combined bilayer -- superconductor system as (see, e.g., Ref.~[\onlinecite{Yokoyama}])
\begin{align}\label{eq:calG}
{\cal G}=(G^{-1}-\hat{T})^{-1}=G\sum_i(\hat{T}G)^i,
\end{align}
where ${\cal G}$ and $G$ are the Green's functions of the whole system, including bilayer and superconductor,
with and without a tunneling operator $\hat{T}$ acting between the bilayer and the superconductor, respectively. Writing $G$ and $\hat{T}$ in matrix form we have
\begin{align}\label{eq:G0Tmatrix}
G=\begin{pmatrix}
G_{SC}&0&0\\0&G^{11}_{BL}&G^{12}_{BL}\\0&G^{21}_{BL}&G^{22}_{BL}
\end{pmatrix}
,\, \,
\hat{T}=\begin{pmatrix}
0&T_{SB}&0\\
T_{BS}&0&0\\
0&0&0
\end{pmatrix}.
\end{align}
Here $G_{SC}$ is the Green's function of the superconductor and $G_{BL}^{ij}$, with $i,j = 1,2$ being layer indices, is the Green's function of the bilayer system, both before any superconducting proximity effect in-between the two.
The coupling between the bilayer and the superconductor is assumed to only be between the superconductor and layer 1 (bottom)  of the bilayer. It is modeled by the tunneling coefficient $T_{SB}=T_{BS}^*= t_0\sigma_0\otimes\eta_z$, where $t_0$ denotes the (single-particle) tunneling parameter and $\eta$ is a Pauli matrix in electron-hole Nambu space.
Note that each component of the $G$ and $T$ matrices are $4\times 4$ matrices in spin and electron-hole Nambu spaces.
Expressed in the electron-hole Nambu space the superconductor has an anomalous pairing term $f_{sc}$ and the Green's function can written in the form
\begin{align}\label{eq:GSC}
G_{SC}=\begin{pmatrix}
g_{sc}&f_{sc}\\ \bar{f}_{sc}& \bar{g}_{sc}
\end{pmatrix},
\end{align}
while the Green's function of the bilayer has no original anomalous part and thus takes the form
\begin{align}\label{eq:HBL}
G_{BL}^{ij}=\begin{pmatrix}
g_{bl}^{ij}&0\\ 0 & \bar{g}^{ij}_{bl}
\end{pmatrix}.
\end{align}
Here $g,\bar{g}$ denotes the normal Green's function in electron and hole spaces, respectively, determined directly by ${\cal \hat{H}}_{SC}$ for the superconductor and  ${\cal \hat{H}}_{BL}$ for the bilayer. The anomalous Green's function for the conventional $s$-wave spin-singlet superconductor in Eq.~\eqref{eq:H SC} takes the form $f_{sc} = -i\Delta_0\sigma_y/(\varepsilon^2-\epsilon_B^2)$, where $\epsilon_B=\sqrt{\epsilon_k^2 + \Delta_0^2}$ is the Bogoliubov quasiparticle energy dispersion of the superconductor.

Using Eq.~\eqref{eq:calG} to calculate ${\cal G}$, the first term adding a contribution to the Green's function of the bilayer system due to the proximity to the superconductor is second order in the tunneling $t_0$:
$G\hat{T}G\hat{T}G$. We are here mainly going to be concerned with the induced anomalous Green's function in the bilayer, which we denote $f_{bl}$. Writing $f_{bl}$ in layer space we arrive at
\begin{align}\label{eq:F-TI}
f_{bl}=|t_0|^2
\begin{pmatrix}
g^{11}_{bl}\, f_{sc}\, \bar{g}^{11}_{bl} \,\,& \, \,
g^{11}_{bl} \,f_{sc}\,\bar{g}^{12}_{bl} \\
g^{21}_{bl} \,f_{sc}\,\bar{g}^{11}_{bl}  \, \, & \, \,
g^{21}_{bl} \,f_{sc}\,\bar{g}^{12}_{bl}
\end{pmatrix}.
\end{align}
From the expression for $f_{bl}$ it is clear that, although we did not assume any direct tunneling between the superconductor and the second layer of the bilayer, superconductivity is induced also into the second layer through the interlayer Green's functions $g^{12}_{bl}$ and $g^{21}_{bl}$, both proportional to $\delta$. Moreover, the off-diagonal terms in Eq.~\eqref{eq:F-TI} describes interlayer pairing, where one electron in the Cooper pair belongs to one layer while the other electron belongs to the other layer.
Using Eq.~\eqref{eq:F-TI} for the anomalous propagator we can calculate the proximity-induced pairing in the first and second layers, as well as the induced interlayer pairing.

%
\section{Results}
\subsection{Energy dispersion and density of states}\label{sec:disp}
Before going into the details of the proximity-induced anomalous Green's function in the bilayer, we first illustrate the effect of the superconductor on the energy dispersion and local density of states (LDOS) of the bilayer. In order to connect to a currently  experimentally studied system,\cite{Wang12} we choose our bilayer to be a TI thin film with no magnetic field, i.e., $K$ and $M$ in Eq.~\eqref{eq:H TI} are both set to zero.
By solving the Bogoliubov-de Gennes equation, $\hat{\cal{H}}_{TOT}\psi=\varepsilon \psi$, for the Hamiltonian $\hat{\cal{H}}_{TOT} =\hat{{\cal H}}_{SC}+\hat{\cal{H}}_{BL}+\hat{T}$, we directly find the energy dispersion of such a TI thin film -- superconductor hybrid structure.
%
\begin{figure}
\includegraphics[width=1.0\linewidth]{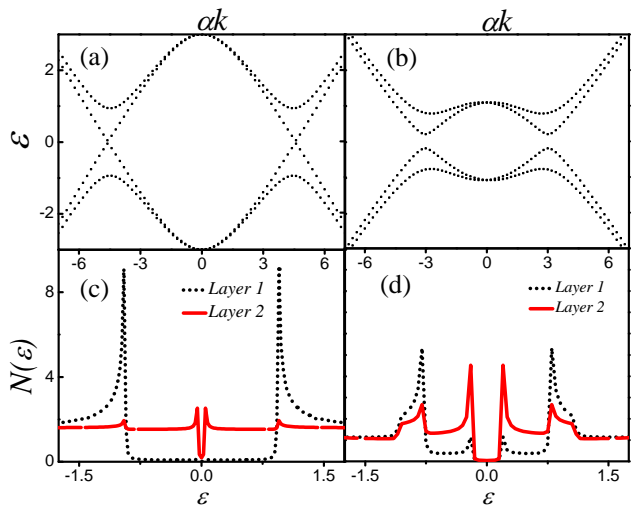}
\caption{(Color online.) (a, b) Energy dispersion (meV) as a function of $\alpha k $ (meV) and (c, d) LDOS ([$\frac{L}{2\pi \alpha}]^2$ states/meV) for each layer of a TI thin film in proximity to an external $s$-wave superconductor. The TI film is described by $K =\mu=M=0$ and $V=5$~meV to model an undoped TI thin film under low bias. The tunneling between the two TI surfaces is $\delta=2$~meV (a, c) and $\delta=4$~meV (b, d). A generic superconductor is assumed by setting $m_{sc}= m_e$ and $\mu_{sc} = 1$~eV, which remove the dispersion of the superconductor from the low-energy regime.
\label{fig:disp-sc}}
\end{figure}
Figures \ref{fig:disp-sc} (a, b) show the dispersion of the TI thin film after proximity to a
superconductor for two different film thicknesses, modeled by different surface hybridization parameters $\delta = 2$~meV (a) and $\delta = 4$~meV (b).
For the smaller surface hybridization only one of the TI surface Dirac points are visibly gapped, and it is only with increasing surface hybridization that both surface Dirac points become clearly gapped, as seen in Fig.~\ref{fig:disp-sc}(b).

To gain more insight into the behavior of each layer, we also calculate the LDOS in each layer, $N_i(\varepsilon)=-1/\pi \sum_k {\rm Tr}[{\rm Im}\{{\cal G}^{ii}_{BL}(\varepsilon+i0^+,k) \}]$. Here, ${\cal G}^{ii}_{BL}$ is the Green's function in the $i$th layer of the bilayer after proximity to the superconductor and the trace is over both spin and electron-hole degrees of freedom.
For the larger surface hybridization, two superconducting gaps with their characteristic coherence peaks are clearly visible, see Fig.~\ref{fig:disp-sc}(d). The smaller gap is mainly associated with layer 2, which is the layer farthest from the superconductor, whereas the larger gap principally belongs to layer 1. The finite DOS within the larger gap is a consequence of the finite surface hybridization, which causes the lowest energy band to not just belong to layer 2 but also have a small component residing in layer 1.
For smaller surface hybridizations, as seen in Fig.~\ref{fig:disp-sc}(c), the superconducting gap in layer 2 is very small, whereas the larger gap, associated with layer 1, does not significantly change in magnitude as a function of $\delta$. The DOS within the larger gap is now also effectively zero since the surface hybridization is very small.

\subsection{Induced anomalous Green's function}\label{sec:Green}
Having established a superconducting proximity effect throughout the bilayer by studying the LDOS, we now focus on the details of the superconducting pairing induced into the bilayer system. For this purpose we carefully examine the anomalous Green's function induced in the bilayer, given in Eq.~\eqref{eq:F-TI}.
In its most general form we can separate the induced anomalous Green's function in the bilayer $f^{ij}_{bl}$ for layers $i,j$ into a spin-singlet part $f_s^{ij}$ and the spin-triplet parts $f_0^{ij}$ for $m_z=0$ and $f_\pm^{ij}$ for $m_z=\pm 1$, where $m_z$ denotes the different $z$-axis projections of the triplet state. Even though the external superconductor is assumed to be a conventional spin-singlet $s$-wave superconductor, we will show in the following that spin-triplet pairing is induced into the bilayer in the presence of finite spin-orbit coupling $\alpha$ or magnetic field $M$.

By using the Hamiltonian in Eq.~\eqref{eq:H TI} to calculate the Green's function of the bilayer system
before proximity to the superconductor and then insert the result into Eq.~\eqref{eq:F-TI}, the induced superconducting anomalous Green's function in the bilayer can be shown to take the form:
\begin{eqnarray}\label{eq:fij}
f^{ij}_\pm &=& \delta^n\Gamma \alpha(k_x\mp ik_y)h_\pm^{ij}(\varepsilon,k^2)\nonumber\\
f^{ij}_0 &=& \delta^n\Gamma M  h_0^{ij}(\varepsilon,k^2)\nonumber\\
f^{ij}_s &=& \delta^n\Gamma h_s^{ij}(\varepsilon,k^2).
\end{eqnarray}
Here, $n$ labels the number of tunneling processes inside the bilayer, with $n = 0,2$ for superconducting pairing within the first and second layers, respectively, i.e.~for intralayer pairing. For interlayer pairing, where the two electrons building up a Cooper pair belongs to different layers, $n = 1$.
The coefficient $\Gamma$ is the same for all pairing terms and is equal to $\Gamma= i\Delta_0|t_0|^2/[(\varepsilon^2-\epsilon_B^2) D^2]$, with
$D=((\varepsilon-K)^2-\xi_+^2)((\varepsilon-K)^2-\xi_-^2)$. Here $\xi_\pm$ is the energy dispersion of the Hamiltonian Eq.~\eqref{eq:H TI} for $K=0$ and it is equal to
$\xi_\pm =\alpha^2k^2+V^2+M^2+\delta^2\pm 2\sqrt{M^2(V^2+\delta^2)+\alpha^2k^2V^2}$. This is the (positive) energy dispersion for a TI thin film displayed in Fig.~\ref{fig:schem}(b).

Even before going into the details of the functions $h^{ij}$ we can discuss the general structure of the induced pairing amplitudes given in Eqs.~\eqref{eq:fij}. First, the existence of $m_z=\pm 1$ spin-triplet pairing is a direct consequence of finite Rashba spin-orbit coupling in the bilayer. Moreover, this pairing term has chiral $p$-wave symmetry due to the $k_x \mp ik_y$ prefactor. The spatial chirality is set directly by the spin quantum number along $z$. The occurrence of spin-triplet chiral $p$-wave pairing for finite spin-orbit interaction is fully consistent with results on proximity-induced superconductivity in TI surfaces.\cite{Fu08, Stanescu10, Black-Schaffer11QSHI, Lababidi11, Black-Schaffer13TISC}
Secondly, the $m_z=0$ spin-triplet state is a consequence of finite magnetization $M$. Interestingly, this pairing term is always even in momentum space, in fact it has $s$-wave symmetry. Due to the fermionic nature of the superconducting wave function this necessitates that intralayer $m_z = 0$ spin-triplet $s$-wave pairing is odd in time, or equivalently frequency (energy), which will be evident in the results following.
This is in agreement with previous results on proximity effect in half-metallic ferromagnets, where spin polarization leads to odd-frequency $m_z=0$ spin-triplet $s$-wave pairing, while spin rotation at the interface leads to equal spin-triplet pairing.\cite{Eschrig} Finally, the spin-singlet part of the induced anomalous Green's function also have $s$-wave symmetry in momentum space. 
Before presenting the full results for a generic bilayer we will in the following discuss experimentally relevant systems in some specific limits.

%
\subsubsection{Single layer}
We first let $\delta\rightarrow 0$ and thus completely decouple the two layers from each other. Since we have assumed that superconductivity only tunnels directly into one of the bilayer layers, the induced anomalous Green's function in the bilayer  then just has a non-zero $f^{11}$ term. This term also significantly simplifies in the $\delta\rightarrow 0$ limit and can be written as:
\begin{eqnarray} \label{eq:single}
f^{11}_{\pm }&=&2\Lambda \alpha(k_x\mp ik_y)(M\mp (V+\mu))\nonumber\\
f^{11}_0&=&2i\Lambda  M \varepsilon \nonumber\\
f^{11}_s&=&-\Lambda(\varepsilon^2+M^2-\alpha^2k^2-(V+\mu)^2),
\end{eqnarray}
where $\Lambda^{-1}$ is equal to $(\varepsilon^2-\epsilon_B^2)D$ and we have for simplicity assumed no parabolic dispersion in the bilayer. For a finite parabolic dispersion, $\mu$ can simply be replaced by $K$. The results in the single layer limit confirm the odd-frequency (odd dependence on the energy $\varepsilon$) $s$-wave nature of $f_0^{11}$. It further shows that the spin-singlet pairing also has $s$-wave symmetry.
Moreover, Eqs.~\eqref{eq:single} show how it is possible, by tuning the external parameters $M, V$, and $\mu$, to manipulate both the amplitude and the very existence of some of the pairing terms. For example,
by setting $M= (V+\mu)$ it is possible to completely eliminate $f_+^{11}$ pairing but keep $f_-^{11}$ pairing.

%
\subsubsection{Undoped TI thin film}
Another case simpler than the most generic bilayer system is that of a TI thin film. Here, $\beta = 0$ in Eq.~\eqref{eq:H TI} and for simplicity we also ignore any possible doping of the system such that $\mu = 0$, and thus $K=0$.
Focusing first on the intralayer spin-triplet pairing we find for the undoped TI thin film
\begin{align}
h^{11}_\pm & =\Lambda_1(V\mp M) \nonumber \\
h^{22}_\pm & =\Lambda_2(V\pm M) \nonumber \\
h^{ii}_0 & = \Lambda_i M \varepsilon 
\end{align}
where $\Lambda_1=\varepsilon^4+(M^2+\alpha^2k^2-V^2)^2-2\varepsilon^2(M^2+\alpha^2k^2+V^2)-\delta^4$ and $\Lambda_2=\varepsilon^2+M^2-\alpha^2k^2-V^2-\delta^2$. Even though the  functional dependences on the external parameters are now more complicated compared to the single layer limit several notable observations are still possible.
Most interestingly, by tuning to the point $V=M$ it is possible to make one of the equal spin components completely disappear in one layer, while the other equal spin component disappears in the other layer. This is a consequence of the different helicities, set by the spin-orbit coupling, in the two surface states. We also again find that the intralayer $m_z = 0$ spin-triplet state is an odd-frequency $s$-wave state.
Keeping in mind that the gap parameter for spin-triplet pairing can also be written in the form $\hat{\Delta}\sim i \textbf{d}(k)\cdot {\bm \sigma}\sigma_y$, we conclude that the $\textbf{d}$-vector takes the form
$\textbf{d}(k)\sim [\alpha(\pm Mk_x-iVk_y), \alpha(-iVk_x \mp Mk_y),\varepsilon M]$ for intralayer pairing in layers 1 and 2, respectively .

The layer degree of freedom also opens for the possibility of interlayer pairing $f^{12}$. More specifically, for each spin and spatial symmetry the interlayer pairing $f^{12}$ can, even in the general bilayer case, be divided up into a part that is even (eL) in layer index and a part that is odd (oL): $f^{12}=f^{12,eL}+f^{12,oL}$. For the part that is odd in layer index this leads directly to the frequency dependence necessarily also being odd, if the spin and spatial symmetries are to be the same as for the intralayer and even-interlayer pairings. This is in analogy with multiband systems where odd-frequency, odd-interband pairing has been shown to have the same spatial and spin symmetries as even-frequency, even-interband pairing.\cite{Black-Schaffer13multi} In a bilayer system, the layer index plays the same role as the band index in a multiband superconductor.

Focusing on the existence of odd-frequency interband pairing, we find the components $h^{12,oL}_\pm=4\Lambda_2 V\varepsilon$, $h^{12,eL}_0=-i4 \Lambda_2 V\varepsilon$, and $h^{12,oL}_s=-\varepsilon \Lambda_2(2V^2+\delta^2)$ for a TI thin film, where the linear energy dependence directly gives the odd-frequency nature. The equal spin pairing is a chiral $p$-wave state (see Eq.~\eqref{eq:fij} for the expression for $f^{12}$) and thus odd-frequency pairing means oddness in layer index. Also the odd-frequency spin-singlet interlayer pairing is odd in layer index since it has $s$-wave spatial symmetry.
On the other hand, the $m_z = 0$ pairing, dependent on a finite magnetization, has $s$-wave symmetry and thus the odd-frequency term is instead associated with even-interlayer pairing.
It should be noted that the dependence on $V$ for the spin-triplet odd-frequency terms is a consequence of zero doping in the TI. At finite doping levels this is replaced by dependence on $\mu \pm V$, as seen in the Appendix \ref{app}.
Also, while we do not explicitly list them here, the corresponding equal spin and spin-singlet even-frequency, even-interlayer pairing and the $m_z = 0$ even-frequency, odd-interlayer pairing are also both present in a TI thin film, with the full analytical expressions given in Appendix \ref{app}.

\subsubsection{General bilayer system}
For a general bilayer system we summarize in Table \ref{tab} all possible pairing amplitudes divided up in terms of their spin quantum numbers and with intra-, even- and odd-interlayer pairing listed separately.
%
\begin{table}[h]
\begin{tabular}{ c||c|c|c }
 \hline \hline
 spin & intralayer & even-interlayer & odd-interlayer\\
 \hline
$f_\pm \sim\alpha$   & $p$-wave, even-$\omega$ & $p$-wave, even-$\omega$ &  $p$-wave, odd-$\omega$\\
$f_0 \sim M$ &  $s$-wave, odd-$\omega$ & $s$-wave, odd-$\omega$  & $s$-wave, even-$\omega$\\
$f_s$ & $s$-wave, even-$\omega$ & $s$-wave, even-$\omega$ & $s$-wave, odd-$\omega$\\
 \hline \hline
\end{tabular}
 \caption{Spatial and frequency ($\omega$) symmetries of the proximity-induced superconducting state in a general bilayer system divided up according to spin quantum numbers and layer properties.}
 \label{tab}
\end{table}
The full analytical forms are given in Appendix \ref{app}.
The overall structure of the table entries is a direct consequence of Eqs.~\eqref{eq:fij}. 
The equal spin-triplet pairing is only present for finite Rashba spin-orbit coupling $\alpha$ and has always chiral $p$-wave symmetry, whereas the $m_z = 0$ spin-triplet pairing requires a finite magnetization $M$ and has $s$-wave symmetry. Due to the spin-triplet $s$-wave nature the latter pairing is necessarily odd in frequency for both intralayer and even-interlayer pairings.
The spin-singlet pairing also has $s$-wave spatial symmetry.

Moreover, intralayer and even-interlayer pairing terms have the same overall behavior. In addition, as discussed above for a TI thin film, odd-interlayer pairing is also generally present. The oddness in layer index is accompanied by a change in the frequency dependence. For spin-singlet and equal-spin pairing, the odd-interlayer pairing is odd in frequency, whereas for $m_z = 0$ spin-triplet pairing the odd-interlayer pairing is even in frequency. Very generally, the symmetry of the superconducting state fulfills the requirement $PTL = 1$ for spin-singlet and $PTL = -1$ for spin-triplet pairing, where $P$ is the spatial parity, $T$ stands for time-reversal, and $L$ is the layer parity.
We note that the interlayer pairing should be possible to detect experimentally using nonlocal spectroscopy of Andreev bound states, as has recently been demonstrated in double quantum dots for inter-dot pairing.\cite{Schindele}

Very generally we can understand the occurrence and parameter dependences of the odd-frequency pairing by the following argument: Odd-frequency pairing requires that the two electrons of the Cooper pair have different energies, such that the overall energy dependence can become odd. This is why intralayer $m_z =0$ spin-triplet pairing can be odd in frequency, because it is proportional to the magnetization and each spin has a different energy in a magnetic field. In the undoped TI thin film limit we can instead keep each layer at a different energy by applying a finite bias $V$. This produces odd-frequency spin-triplet  pairing in-between the layers, i.e. interlayer pairing, proportional to $V$. However, there in fact already exists another asymmetry between the layers in a bilayer -- superconductor hybrid structure, since only one of the layers is in direct proximity to the superconductor. This is why odd-frequency interband pairing appears very generally for all bilayer systems, and not just for a limited range of parameters.

Beyond the general symmetries displayed in Table \ref{tab}, the $f^{ij}$ terms also have additional functionality of the input parameters $\alpha, K,\delta,V,M,\mu$. There exists multiple special cases where tuning these input parameters can result in an engineered superconducting state.
For example, as already mentioned, in an undoped TI thin film setting $M = V$ eliminates $f^{11}_+$ and $f^{22}_-$, such that the resulting equal spin state only has $k_x + ik_y$-wave symmetry in layer 1 and $k_x-ik_y$-wave symmetry in layer 2. There is thus a fully chiral superconducting state in each layer for $M = V$.
This difference arises from the opposite spin-orbit coupling direction in the two different TI surfaces. At finite doping levels $f^{22}_-$ still vanishes but $f^{11}_+$ is now linearly dependent on $\mu$.
Another experimentally relevant tuning is when the Rashba coefficient $\alpha \rightarrow 0$, which describes a double quantum well system with no spin-orbit coupling. \cite{Hasbun03} For no magnetic field all spin-triplet components are then zero and only spin-singlet superconductivity persists. A finite magnetization $M$ generates $s$-wave spin-triplet $m_z = 0$ pairing, with odd-frequency dependence.

%
\subsection{Superconducting order parameter}\label{sec:OP}
Having calculated the proximity-induced anomalous Green's function it is also possible to extract the order parameters for the different pairing symmetries given in Eq.~\eqref{eq:fij}.\footnote{Technically these are pairing amplitudes, since superconductivity is proximity-induced into the bilayer and thus not generated by an internal pair potential.} The momentum-dependent order parameters are given by (see, e.g.,~Ref.~[\onlinecite{Continentino}])
\begin{equation}\label{eq:delk}
\Delta(k)=\int_{-\infty}^\mu {\rm Im}\{f(k,\varepsilon)\} d\varepsilon.
\end{equation}
Alternatively, we can also obtain overall amplitudes of the order parameters by integrating over all possible momenta. For $p$-wave symmetry we, however, then need to exclude the phase $e^{i\phi_k}$, $\phi_k=\tan^{-1}(k_x/k_y)$, when we determine the overall amplitude of the order parameter.

\begin{figure}
\includegraphics[width=1.0\linewidth]{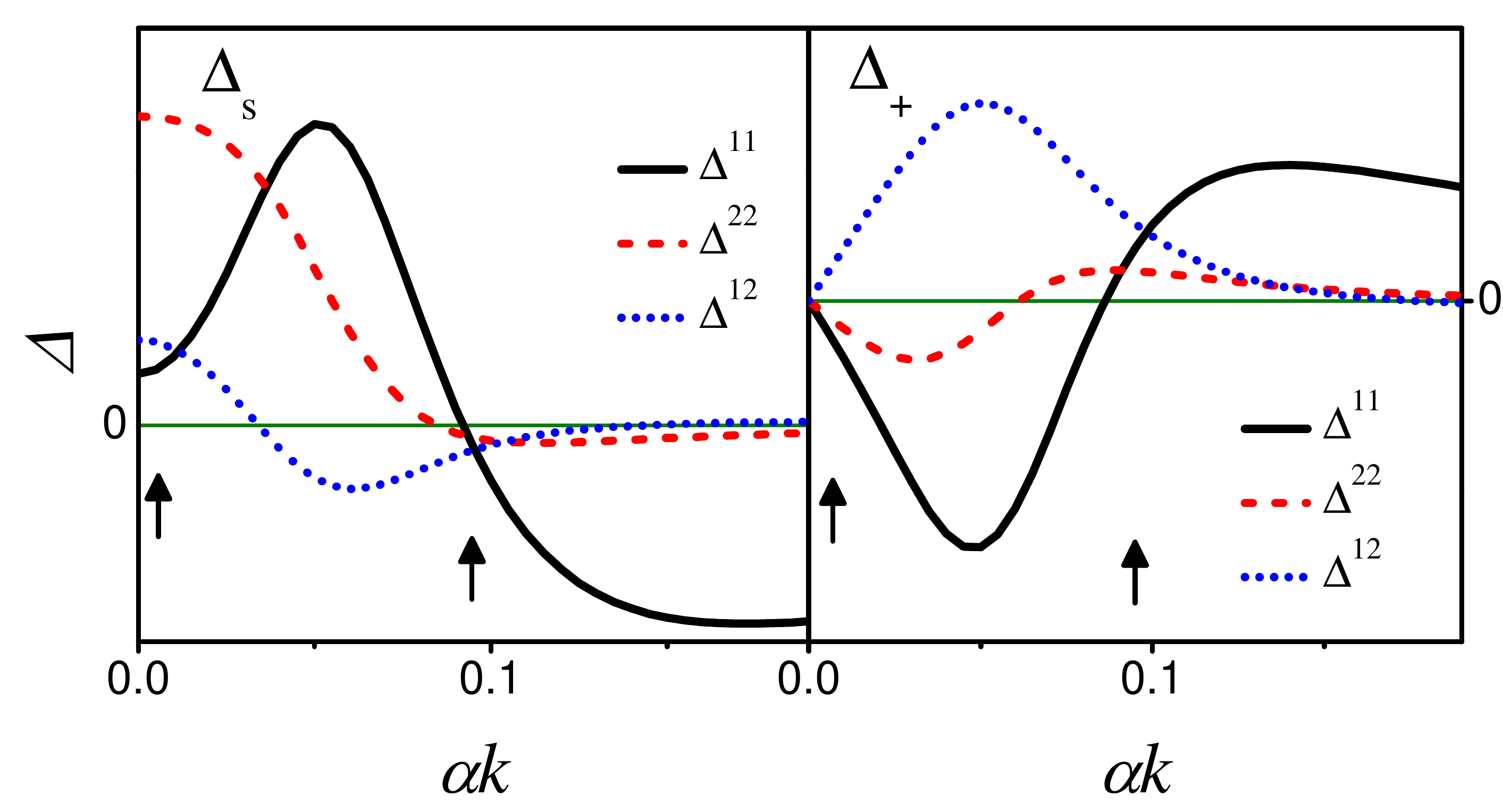}
\caption{(Color online.) Momentum-dependent order parameters $\Delta^{11}$ (intralayer pairing in layer 1, solid black), $\Delta^{22}$ (intralayer pairing in layer 2, dashed red), and $\Delta^{12}$ (interlayer pairing, dotted blue) as a function of momentum scaled with the spin-orbit coupling constant $\alpha$ (eV) for spin-singlet (left) and spin-up spin-triplet (right) pairing. Additional parameters are $V=0.05$, $\delta=0.05$, and $\mu=0.07$~eV. Arrows indicate the position of the two different Fermi momenta, green thin line indicates the zero level.
\label{fig:Deltak}}
\end{figure}

In order to limit the possible parameter space, but still keep an experimentally relevant bilayer system in mind, we here focus on TI thin films at finite doping levels and with applied electric bias. 
Figure \ref{fig:Deltak} shows $\Delta(k)$ for spin-singlet (left panel) and spin-triplet $m_z=+1$
(right panel) pairing for such a TI thin film. For the chosen doping level the Fermi level is located towards the bottom of the conduction band. The combined effect of Rashba spin-orbit coupling and applied bias potential between the layers is that the band dispersion has a Rashba splitting and as a result an inner and an outer Fermi surface appear, as illustrated in Fig.~\ref{fig:schem}(b). The two different Fermi momenta are indicated by arrows in Fig.~\ref{fig:Deltak}.
%
As clearly seen in the figure, all pairing terms changes signs as function of the momentum. Even when we only focus on the sign of the order parameters at the two different Fermi momenta, most of them still changes signs between the two Fermi surfaces.
Incorporating this sign change in the order parameter symmetry, the spin-singlet state should properly be addressed as a $s_\pm$-wave symmetry, whereas the spin-triplet state becomes a $(p_x+ip_y)_\pm$ state, where the sign subscript indicates a sign change between the two Fermi surfaces. Such a sign-changing $s$-wave state is currently most prominently known from the iron-pnictide superconductors,\cite{Mazin10} although there the order parameter switch sign between the $\Gamma$ and $M$ points, whereas in the present case the sign change is between Fermi surfaces both centered around $\Gamma$.
Spin-orbit coupled semiconductors in proximity to a $s_\pm$-wave superconductor, such that the inner and outer Fermi surfaces have effective order parameters with opposite signs, have recently been proposed as a route to engineer time-reversal invariant topological superconductors, which hosts pairs of Majorana boundary states.\cite{Deng12, Zhang13} Here we find that for TI thin films the necessary sign switch in the superconducting order parameter between the two Fermi surfaces is present even for a conventional $s$-wave superconductor, simply due to the bilayer structure.
Note that we used $M=0$ here in order to explicitly show that a time-reversal invariant non-trivial superconducting state is in generally present in biased TI thin films.

\begin{figure}
\includegraphics[width=1.0\linewidth]{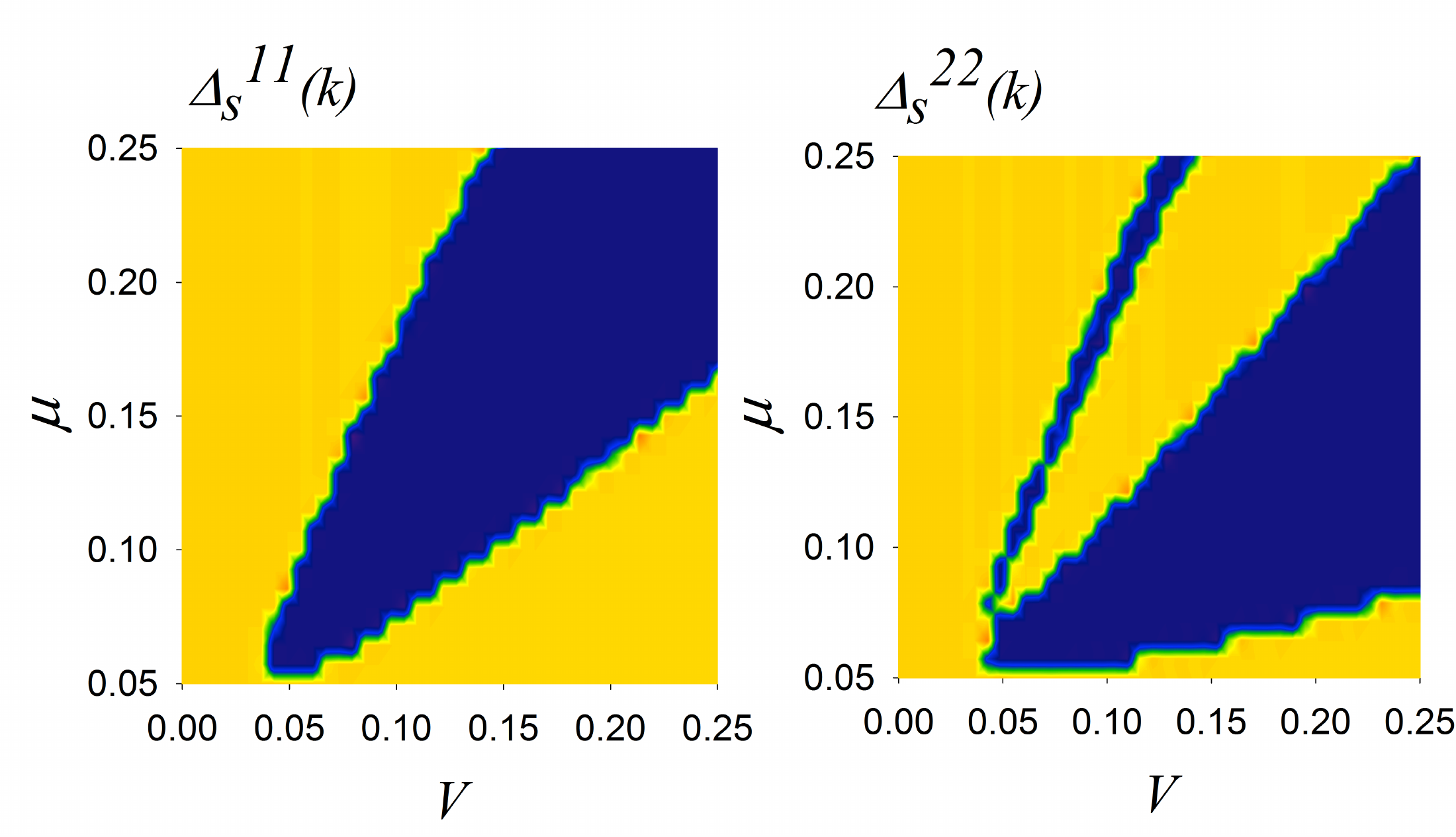}
\caption{(Color online.) Contour plots showing the parameter region in $V-\mu$ (eV) space where the spin-singlet order parameter changes sign between the two Fermi surfaces $\Delta_s^{ii}(k_{F1})\Delta_s^{ii}(k_{F2})<0$ (blue) for a TI thin film, demonstrating $s_\pm$-wave pairing in layer (surface) 1 (left figure) and 2 (right figure). The film thickness is represented by $\delta=0.05$~eV.
\label{fig:Sign}}
\end{figure}

To further explore the sign switching between the two Fermi surfaces we plot ${\rm sign}(\Delta_s^{ii}(k_{F1})\Delta_s^{ii}(k_{F2}))$ in $V-\mu$ space for both TI surfaces in Fig.~\ref{fig:Sign}.
The blue (darker) color shows the parameter space where the intralayer spin-singlet order parameter $\Delta_s$ switches its sign between the two Fermi surfaces. The somewhat different behavior of the layers 1 and 2 can be traced back to the different expressions for proximity-induced superconductivity in Eq.~\eqref{eq:F-TI}. In layer 1 superconductivity is a consequence of a direct tunneling process from the superconductor and given by $g_{11}f_s\bar{g}_{11}$, whereas in layer 2 superconductivity is also dependent on the interlayer (intersurface) tunneling process and expressed through $g_{21}f_s\bar{g}_{12}$.

\begin{figure}
\includegraphics[width=1.0\linewidth]{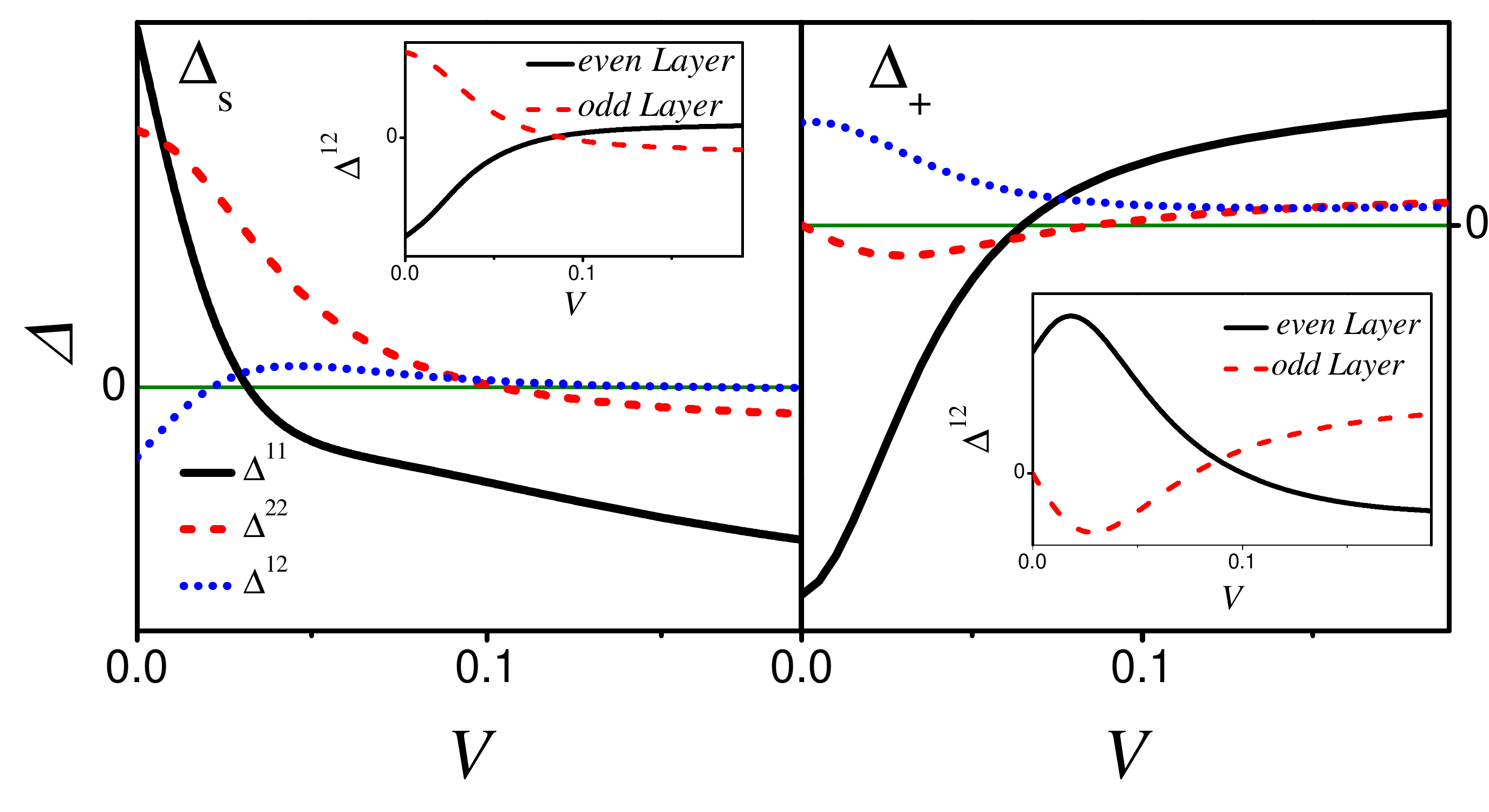}
\caption{(Color online.) Momentum-integrated order parameters $\Delta^{11}$ (intralayer pairing in layer 1, solid black), $\Delta^{22}$ (intralayer pairing in layer 2, dashed red), and $\Delta^{12}$(interlayer pairing, dotted blue) as a function of applied bias field $V$ (eV). Additional parameters are $\delta=0.05$~eV, $\mu=0.1$~eV. The insets show the even and odd-interlayer pairing terms. 
\label{fig:DeltaV}}
\end{figure}

Beyond a sign change for the momentum-resolved order parameter $\Delta(k)$, there can also be a sign change for the order parameter between the two layers in a bilayer system. In Fig.~\ref{fig:DeltaV} we show intralayer and interlayer momentum-integrated pairing plotted as a function of electric bias between the two layers.
As seen, for a range of bias potentials it is possible to have $\Delta^{11}\Delta^{22}<0$ for both the spin-singlet $s$-wave and the spin-triplet chiral $p$-wave paring. This means that the order parameters change sign in the region between the two layers, creating an effective $\pi$-junction between the two TI surfaces. The $\pi$ phase difference in the intralayer order parameters between the two different layers is a consequence of the asymmetry between the two layers.
Interestingly, a two-wire system in proximity to an $s$-wave superconductor has very recently been shown to host a time-reversal invariant topological superconducting state if the two wires have different Rashba spin-orbit couplings and the determinant of the pairing matrix in wire space is negative.\cite{Keselman13, Flensberg14} In the parameter space where $\Delta^{11}\Delta^{22}<0$ for a TI thin film these two conditions are easily satisfied between the two surfaces. A bilayer system with only one layer in proximity to a conventional $s$-wave superconductor might thus represent a 2D version of the same mechanism, only requiring tuning of an external electric bias potential.

%
\section{Concluding remarks}\label{sec:sum}
We have in this work studied bilayer -- superconductor hybrid structures, where a conventional spin-singlet $s$-wave superconductor is in direct proximity contact with one of the layers in the bilayer. In order to address general bilayer systems we have used a model for a bilayer which includes general parabolic dispersion, Rashba spin-orbit coupling, layer tunneling, bias potential, and magnetic Zeeman field.

By extracting the proximity-induced anomalous Green's function in the bilayer we have been able to show how, and also what type of, superconductivity is proximity-induced in the bilayer system. Due to the layeredness, possible spin-orbit coupling, and tuning by electric and magnetic fields, a very rich structure of the superconducting pairing can appear in a bilayer. Even though we always assume that the external superconductor has spin-singlet $s$-wave symmetry, equal-spin spin-triplet $p_x\pm ip_y$-wave pairing is induced in both layers in the presence of a finite spin-orbit coupling, and $m_z = 0$ spin-triplet $s$-wave pairing with an odd dependence on frequency (energy) appears whenever a magnetic Zeeman field is applied.
Moreover, not only intralayer pairing but also interlayer pairing, where the two electrons building up a Cooper pair belong to different layers, is present in a bilayer. The interlayer pairing can be either even or odd in the layer index, and this additional index opens the possibility for additional odd-frequency pairing. More specifically, we find a complete reciprocity between oddness in frequency and in layer index. For example, spin-singlet $s$-wave pairing exists both as even-frequency intralayer (and even-interlayer) pairing as well as odd-frequency, odd-interlayer pairing. This is analogous to the situation recently discovered in multiband superconductors, where odd-frequency, odd-interband pairing can be generated.\cite{Black-Schaffer13multi} Note, however, that interlayer pairing is naturally present in bilayer systems such as TI thin films, whereas interband pairing in a multiband superconductor requires additional band hybridization mechanisms.

We have also calculated the induced superconducting order parameters in doped TI thin films, as a contemporary example of a bilayer system. Such films have both an inner and outer Fermi surface in the presence of a finite bias voltage, and we find that over a wide range of parameters the order parameters switches signs between the two Fermi surfaces as well as between the two layers (surfaces).
Interestingly, order parameter sign-switches, both in reciprocal space between Fermi surfaces\cite{Deng12,Zhang13} and in real space,\cite{Keselman13,Flensberg14} have recently been proposed as a key to produce time-reversal-invariant topological superconductors with pairs of Majorana fermions at surfaces. This opens up for the possibility of using TI thin films in proximity to conventional $s$-wave superconductors to engineer topological superconductors with Majorana fermions and is the subject of future work.

Beyond the structure and symmetry of the superconducting state we have also reported on the energy dispersion and LDOS in the bilayer system, with superconducting energy gaps clearly present in both layers at finite interlayer hybridization.
In reference to the presence of odd-frequency superconductivity we point out that we find no evidence of zero- or low-energy states in the superconducting gap. Zero-energy and sub-gap states have been associated with odd-frequency pairing appearing at surface and interfaces,\cite{Tanaka07, Tanaka07PRB,Yokoyama07,Tanaka12,Asano13} and also under more general terms.\cite{Balatsky92, Dahal09} The lack of low-energy states in bilayer systems is, however, in agreement with other recent findings of odd-frequency pairing in multiband superconductors,\cite{Black-Schaffer13TISC, Black-Schaffer13multi} in spatially inhomogenous TIs,\cite{Black-Schaffer12oddw} as well as in heavy-fermion compounds.\cite{Coleman93}

Let us also briefly comment on the range of values for the input parameters in our calculations in terms of experimental accessibility. Focusing mainly on the experimental parameters for a TI thin film, a prototypical spin-orbit coupled bilayer system, the hybridization tunneling $\delta$ between the two surfaces of the TI thin film can be tuned by changing the thickness of the sample from $0$ for samples thicker than six quintuple layers to $0.25$~eV for films only two quintuple layers thick.\cite{Zhang10}
The chemical potential $\mu$ and the potential difference between the two layers $V$ can be tuned electrically by a gate voltage. A range of at least a few hundred meV has been reported.\cite{Zhang10}
Magnetic fields up to 1 T, proportional to the order of $0.2$~meV in magnetic energy $M$, has also been used experimentally,\cite{Wang12,Chang13} although higher magnetic fields should reasonably also be accessible. 
In terms of the external superconductor, the order parameter is of the order of a few meV for conventional $s$-wave superconductors.\cite{Wang12} The tunneling $t_0$ between the superconductor and the bottom layer of the bilayer system is harder to estimate but is expected to be small, due to interface imperfections between the bilayer and the superconductor. Practically, the tunneling amplitude is often tuned in such as way as to reproduce experimental data.
We have in this work explicitly chosen the input parameters in all figures to be well within this current experimental range (apart from the superconducting gap in Fig.~2, which for illustrative purposes was chosen to be large), such that our results should  be  experimentally accessible using common bilayer and superconductor materials.

In summary, a bilayer -- superconductor hybrid structure provides a very rich playground for multiple types of proximity-induced superconductivity, including different spin states and odd-frequency pairing. The bilayer structure also offers the possibility of sign switches of the order parameter, both between the multiple Fermi surfaces and between the two layers, with the possibility of generating Majorana fermion pairs.

\acknowledgments
We thank A.~V.~Balatsky and J.~Fransson for valuable discussions and acknowledge financial support from the Swedish Research Council (Vetenskapsr\aa det, VR) and the G\"{o}ran Gustafsson foundation.

\appendix
\section{Analytic form of the induced anomalous Green's function}\label{app}
In this appendix we give the analytical form of the functions $h^{ij}$ introduced in Eq.~\eqref{eq:fij}. The results are directly applicable to a TI thin film with finite doping, applied bias field, and magnetic field. Replacing $\mu$ in the equations below with $K=\beta k^2 +\mu$ also gives the results for a bilayer with finite parabolic dispersion.

First, for the layer in direct proximity contact with the superconductor (layer 1) we find the intralayer terms:
\begin{widetext}
\begin{eqnarray}\label{eq:h11}
h^{11}_\pm(\varepsilon,k^2)&=&-2[(V\mp M)(\mathfrak{L}^2-\delta^4-4\varepsilon^2(V-\mu)^2)+\mu((\mathfrak{L}+\delta^2)^2-4\varepsilon^2(V-\mu)^2)]\nonumber\\
h^{11}_0(\varepsilon,k^2)&=&2 i \varepsilon[\mathfrak{L}^2-\delta^4-4(V-\mu)(\varepsilon^2(V-\mu)+\delta^2\mu)]\nonumber\\
h^{11}_s(\varepsilon,k^2)&=& (V^2-M^2)[(\mathfrak{L}-\delta^2)^2-4\varepsilon^2(V-\mu)^2)]
+(\alpha^2k^2-\varepsilon^2+\mu^2)[(\mathfrak{L}+\delta^2)^2-4\varepsilon^2(V-\mu)^2)]\nonumber\\
&&+2V[\mu(\mathfrak{L}^2-\delta^4-4\varepsilon^2(V-\mu)^2)-4\varepsilon^2\delta^2(V-\mu)].
\end{eqnarray}
\end{widetext}
Here we have introduced $\mathfrak{L}=M^2+\alpha^2k^2-(V-\mu)^2-\varepsilon^2$, which is an even function of momentum and energy. 
For intralayer pairing in layer 2 we get:
\begin{eqnarray}\label{eq:h22}
h^{22}_\pm(\varepsilon,k^2)&=&4(V\pm M)(\mathfrak{D}\mp2M\mu)\nonumber\\
h^{22}_0(\varepsilon,k^2)&=&-4 i \varepsilon (\mathfrak{D}-2\mu^2) \nonumber\\
h^{22}_s(\varepsilon,k^2)&=& \mathfrak{D}^2+4(\alpha^2k^2V^2+\varepsilon^2(M^2-\mu^2)\nonumber\\&&-M^2(\alpha^2k^2+\mu^2)).
\end{eqnarray}
Here we define $\mathfrak{D}=\varepsilon^2+\mu^2+M^2-\delta^2-V^2-\alpha^2k^2$, which is also an even function of $k$ and $\varepsilon$. This results in the functions $f^{ii}_ \pm$ and $f^{ii}_s$ being even in frequency (energy, $\varepsilon$), while $f^{ii}_0$ is odd in frequency. 

For interlayer pairing the layer degree of freedom is also important and we separate the results into pairing that is even or odd in the layer index, i.e.~\emph{i.e.} $L=\pm 1$.
For even-interlayer ($eL$) pairing we arrive at:
\begin{widetext}
\begin{eqnarray}
h^{12,eL}_\pm&=&4\varepsilon^2(V^2-\mu^2)-\delta^2[\mathfrak{D}-2(M^2\pm2M\mu+V(\mu-V))]
+\mathfrak{L}[\mathfrak{D}+2(M^2\mp2M\mu-V(\mu+V))]\nonumber\\
h^{12,eL}_0&=&4i(\mu-V)\varepsilon[\mathfrak{D}-2\mu^2]\nonumber\\
h^{12,eL}_s&=&2\varepsilon^2(\mu-V)[\mathfrak{D}+2\mu^2+2V\mu-2M^2]+(\mathfrak{L}-\delta^2)
[\mathfrak{D}V-2M^2\mu]
+(\mathfrak{L}+\delta^2)
[\mathfrak{D}\mu-2\alpha^2k^2V-2\varepsilon^2\mu].
\end{eqnarray}
\end{widetext}
With the layer dependence being even, the frequency symmetry is the same as before, i.e.,~$h^{12,eL}_\pm$ and $h^{12,eL}_s$ are both even in frequency, while $h^{12,eL}_0$ is odd. 
For odd-interlayer ($oL$) pairing we find:
\begin{widetext}
\begin{eqnarray}
h^{12,oL}_\pm&=&-2\varepsilon[(\mu-V)(\mathfrak{D}+2M^2\mp4M\mu)+(\mu+V)(\delta^2-2V(\mu-V))]
\nonumber\\
h^{12,oL}_0&=&-i[2(\mathfrak{L}+\delta^2)(\alpha^2k^2+\mu^2-\varepsilon^2)
-(\mathfrak{L}-\delta^2)(\mathfrak{D}-2\mu V)-4\varepsilon^2(V-\mu)^2]\nonumber\\
h^{12,oL}_s&=&-\varepsilon[4\varepsilon^2(\mu-V)\mu+\mathfrak{D}(2V^2+\delta^2-2\mu^2)].
\end{eqnarray}
\end{widetext}
Here the situation is thus reversed compared to the even-interlayer case with $h^{12,oL}_\pm$ and $h^{12,oL}_s$ being odd in frequency, while $h^{12,oL}_0$ has an even-frequency dependence.


\end{document}